 \documentclass[twocolumn,secnumarabic,amssymb, nobibnotes, aps, prd]{revtex4-1}
   \usepackage{graphicx}      
   
\begin{document}   \title{A heuristic rule for binary superlattice co-assembly: Mixed plastic mesophases of hard polyhedral nanoparticles}    \author{Mihir R. Khadilkar}%  
         \affiliation{Department of Physics, Cornell University}\textbf{}  
           \author{Fernando A. Escobedo}%\textbf{}   
            \email[ ]{fe13@cornell.edu}
             \affiliation{Department of Chemical and Biomolecular Engineering, Cornell University} 
             \date{\today}%    
              \begin{abstract} Sought-after ordered structures of mixtures of hard anisotropic nanoparticles can often be thermodynamically unfavorable due to the components' geometric incompatibility to densely pack into regular lattices. A simple compatibilization rule is identified wherein the particle sizes are chosen such that the order-disorder transition pressures of the pure components match (and the entropies of the ordered phases are similar). Using this rule with representative polyhedra from the truncated-cube family that form pure-component plastic-crystals, Monte Carlo simulations show the formation of plastic-solid solutions for all compositions and for a wide range of volume fractions.  \end{abstract}     \maketitle 
              %\tableofcontents     
                 Polyhedral colloidal nanoparticles are versatile building blocks towards designing novel materials with targeted emergent properties. Recent developments in experimental techniques\cite{Henzie2012,Evers2013,Seo2006,Compton2007,Niu2010,Long2011,Shevchenko2006} to controllably synthesize and manipulate polyhedral nanoparticles have fueled many theoretical \cite{Gabrielli2012,Conway2011} and simulation studies\cite{Agarwal2011,Smallenburg2012,Gantapara2013,Torquato2009,Torquato2009PRE,Jiao2009,Marechal2013,Damasceno2012,Damasceno2012Science,Haji-akbari2012,Agarwal2012}  to understand their packing and phase behavior. These building blocks have been shown to exhibit a rich phase behavior at finite osmotic pressures  unveiling the presence of novel mesophases. A mesophase is a partially ordered phase whose properties are intermediate between those of disordered liquids and ordered crystals, such as liquid-crystals, rotator plastic-crystals,  and quasicrystals.   \par Binary mixtures of polyhedra\cite{Khadilkar2013} exhibit a competition between mixing and packing entropy that often leads to phase separation at high pressures; indeed, assembly into binary superlattices using just entropic forces is difficult to achieve \cite{Khadilkar2012}. An earlier study\cite{Khadilkar2013} on the miscibility trends of binary polyhedra mixtures revealed the importance of  the relative size ratio of the components and  of similarity in their mesophase behavior\cite{Agarwal2011}. One of our aims  is to identify shapes and sizes that favor the formation of entropic rotator mixtures.       \par  A  family of truncated cubes, which is readily synthesizable \cite{Seo2006,Compton2007}, has been recently shown to exhibit a diverse set of phases\cite{Gantapara2013}. Further, the kinetics of the disorder-to-order transition for some members of this family has been shown to be substantially faster than that of hard spheres\cite{Thapar2013}, making them appealing choices for applications requiring fast self-assembly. In addition to cuboctahedra (COs) and truncated octahedra (TOs), we choose here a truncated cube with truncation parameter 0.4 (TC4)  \cite{Gantapara2013}, since, like COs and TOs, TC4 also exhibits a rotator mesophase\cite{Gantapara2013}. These choices are motivated by the hypothesis that mesophasic partial disorder can provide enough structural leeway to facilitate ordered solutions to form despite the entropy costs associated with differences in packing. The {\em main} mixtures studied are the three possible pairings  of these three shapes, and are denoted henceforth as COTO, TC4TO and TC4CO.              \begin{widetext}         \begin{figure}  \centering          \includegraphics[height=13cm]{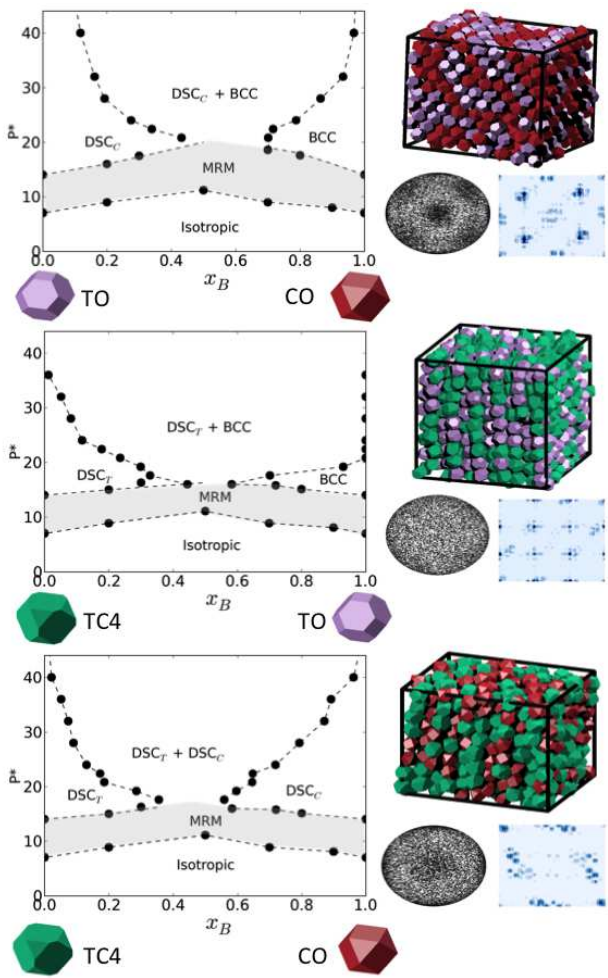}           \caption{Pressure (P*) vs. composition ($x_B$) phase diagram for the 3 main mixtures. DSC$_C$ and DSC$_T$ are distorted simple cubic structures of COs and TC4s respectively\cite{Gantapara2013,Torquato2009PRE}. $x_B$ represents fraction of COs in the COTO and TC4CO mixtures, and the fraction of TOs in the TC4TO mixture. Each diagram is accompanied by a snapshot of the mixed rotator mesophase (MRM) for $x_B$=0.5 (at P*=11.2, 9.6 and 9.6 for COTO, TC4TO and TC4CO respectively), its orientational correlation plot and diffraction pattern. }  \label{fgr:mainfig}  \end{figure}  \end{widetext}                   \par For any target solid mixture, the relative component size-ratio is an important determinant to control the crystal lattice spacing. A recent study\cite{Khadilkar2013} suggested that the solid miscibility in a binary mixture of polyhedra can be linked to the relative values of the order-disorder transition pressure or ODP.  In that study \cite{Khadilkar2013}, however, the components' ODPs were always substantially different and very limited solid miscibility was observed; hence, the questions of what happens when the ODPs matched and whether that provides optimized mesophase compatibility were left  open. For  the present simulations, we set the relative particle size ratios such that their ODPs are approximately equal, which coincidentally entail near-equal circumradii; namely the  ratios of circumradii are CO:TO = 1:1 for COTO, TC4:TO = 1.01:1 for TC4TO and TC4:CO = 1.01:1 for TC4CO (for TC4  we use the largest circumscribing radius).  While equal circumradii is an equivalent criterion to $\Delta$ODP=0 for the main mixtures considered here, we will also use a fourth mixture of spheres and cubes to show that $\Delta$ODP=0 optimizes the {\it overall} miscibility even when equal circumradii does not.      \par For the main mixtures, we probed the phase behavior as a function of pressure using hard-particle Monte Carlo simulations in the isothermal-isobaric ensemble,  including swap moves between the position of particles of different species\cite{Khadilkar2013}. We used interfacial runs to test the relative stability of the phases near a phase transition. While most simulations used equimolar mixtures, additional runs for other compositions were used to more completely map out the phase diagram. Orientational order was analyzed by using the P$_4$ order parameter \cite{John2008} and orientational scatterplots \cite{SM}, while the translational order was probed by using Steinhardt's order parameters Q$_4 $ and Q$_6$\cite{SteinhardtNelson1983} and diffraction patterns (structure factors). To further characterize positional order, we also identified the contributions of FCC, BCC or HCP-like motifs \cite{Agarwal2011} by calculating the distributions of two local bond order parameters ($\bar{q_4}$ and $\bar{q_6}$) (see details in the supplemental material \cite{SM}).        \par The COTO, TC4TO, and TC4CO mixtures exhibit a mixed rotator mesophase (MRM) in between the isotropic phase at low pressures and a phase separated state with two crystalline phases at high pressures (see Figure \ref{fgr:mainfig}). This MRM is stable for all compositions in all three mixtures and for a sizable range of volume fractions\cite{SM}. It is of interest to characterize such novel MRM since the rotator phases of the pure components are distinct in both translational order and rotational disorder. For instance, after the ODP the COs and TC4s rotator phases transform into the orientationally ordered crystal via a first-order transition at the mesophase-to-crystal transition pressure \cite{Gantapara2013}; in contrast, TOs transform continuously into a crystal phase \cite{Thapar}. Below we examine the  properties of the MRM  giving representative results for the COTO mixture.       \par
\begin{figure}    \centering   \includegraphics[height=8cm]{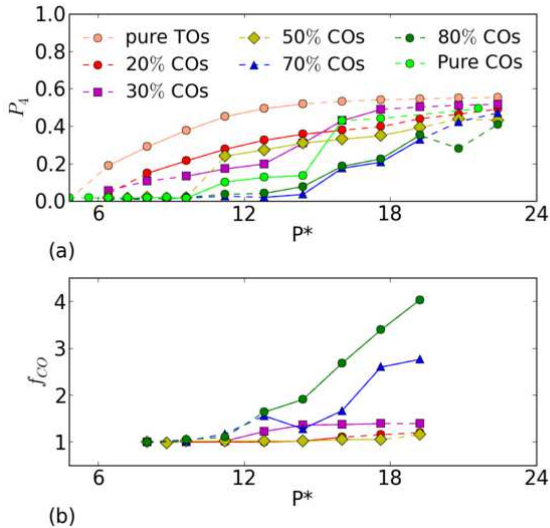}      \caption{Plots showing the effect of changing the mesophase composition in COTO mixture (solid part of each curve represents the stable MRM region). (a) Variation of P$_4$   as a function of pressure, P* . (b) Pressure dependence of the enrichment factor.  }      \label{fgr:cotoxplots}        \end{figure}

  In a purely entropic scenario,  mixtures (that do not form tessellating compounds\cite{Khadilkar2012}) would be expected to phase separate at high pressures into nearly pure component solids to allow denser packings. For our  ODP-matched mixtures,  the MRM delays the onset of phase separation (e.g.,  P* $\approx 21$ in the equimolar COTO). The observed MRM has intermediate orientational order ( P$_4$) as shown in Figure \ref{fgr:cotoxplots}-a for the COTO mixture, and strong positional order (Q$_4$ and Q$_6$). Local compositional heterogeneity or incipient `clustering' can be detected by the average fraction of like-shaped nearest neighbors to a given particle. This fraction should equal the overall composition of the given species in the bulk for an ideal mixture, but it will exceed that as clustering and a tendency for phase separation ensues. We observe that for all three mixtures  the ratio of  local to global composition or `enrichment factor' ($f$) steadily increases with pressure from its ideal (well-mixed) value until eventually reaching the solid-solid phase separated state (Figure \ref{fgr:cotoxplots}-b). The  more symmetric compositions have larger ideal mixing entropy and hence enrichment factors closer to unity. For some of the more skewed compositions, the MRM crystallizes before phase-separating as pressure increases. Figure \ref{fgr:cotoxplots}-a shows how the mesophase-to-crystal transition (as detected by the approach of P$_4$ to the threshold value of ~0.4 for orientional order) changes from being nearly continuous for low CO-compositions (similar to pure TOs) to having more abrupt increases for higher CO-compositions (like pure COs \cite{Thapar2013}).       \par Given that none of the MRMs simulated had one of the known perfect lattice structures, we  obtained the {\it fractions} of different standard structural motifs in the simulated configurations \cite{Agarwal2011}. We observe that  in the equimolar MRMs containing TOs (COTO and TC4TO), the fraction of BCC (which is the target structure for TOs, the better-packing shape in the mixture) {\it increases} with volume fraction (see \cite{SM}). Similarly, for TC4CO MRM,  the fraction of HCP (which is closer to DSC$_C$ and DSC$_T$ structures that COs and TC4s favor respectively) increases with volume fraction.                  %\begin{widetext}    
                    \begin{table}[h] \small  \caption{Summary of results for the miscibility range for COTO mixture in the original and changed size ratios $V_l$ and $V_s$ correspond to the volume of the larger and smaller particle in the mixture respectively. $E_l$ and $E_s$ denote edge lengths while $R_l$ and $R_s$ denote circumradii.}  \label{tbl:COTOdata}  \begin{tabular*}{0.5\textwidth}{@{\extracolsep{\fill}}llllllll}   \hline  System & $\Delta$ ODP & $\Delta P^*_m$ &$\Delta \phi_m$&$A_{MRM}$ &$V_l /V_s$ & $E_l /E_s$& $R_l/R_s$ \\   \hline   O    &$ \approx$ 0.0 &  13 &0.17&7.0&1.21 & 1.58&1.0 \\   S & $\approx$ +1.1& 9.0 &0.13&6.1&1.04&1.66&1.05\\   L  & $\approx$ -1.0 & 3.0 &0.09&3.6&1.41 & 1.50&1.05\\        \hline  \end{tabular*} \end{table}    
                       %\end{widetext}      
                          \par To test whether the equal ODP rule maximizes rotator miscibility, we use the COTO mixture as testbed and change $\pm 5\%$ the relative size ratio by slightly perturbing the size of TOs from its original value (assumed unity,  system O), to be 1.05 ( system L, for larger TOs) and 0.95 (system S, for smaller TOs). This rescaled the ODP of the corresponding TOs from 7.1 (system O) to $7.1\times0.95^3=$ 6.1 (system L) and $7.1\times1.05^3=$ 8.2 (system S). The first observation is that systems L and S also exhibit an MRM over the whole range of compositions, showing that this MRM behavior is robust to small changes of particle size ratios (e.g., size polydispersity that may arise from the experimental synthesis). The extent of miscibility in the MRM can be quantified by using several metrics, e.g.: (1) $\Delta P^*_m$: The difference between the highest and lowest pressure where the equimolar MRM phase is stable, (2) $\Delta \phi_m$: The difference between the highest and lowest volume fraction where the equimolar MRM phase is stable, and (3) $A_{MRM}$: The area where the MRM exists in the volume fraction vs. composition phase diagram. We observed that the extent of miscibility as inferred from all metrics  {\it decreased} \textbf for systems L and S relative to system O. Further, in a previous study\cite{Khadilkar2013} where the size ratio was 63 $\%$  the ODP-matching value, no MRM formed for a wide range of compositions.   \par While the CO:TO volume ratio is not a good predictor of MRM miscibility as it is closer to unity in the L case than in the O case (see Table \ref{tbl:COTOdata}), the ratio of circumradii is. Equal-circumradii, which also holds for the TC4CO and TC4TO mixtures described earlier, could be envisioned as allowing two low-asphericity polyhedral components to freely rotate, effectively sweeping equal spherical volumes in the lattice sites of the MRM. This picture is too simplistic, however, since TOs do not freely rotate in their mesophase\cite{Thapar2013}.    \par
\begin{figure}  \centering   \includegraphics[height=10cm]{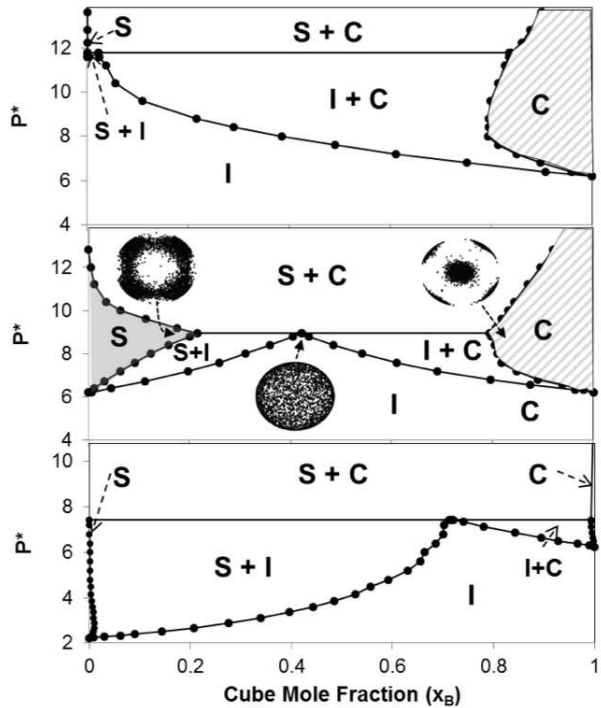}     \caption{Pressure-composition phase diagrams for spheres (diameter $\sigma$) and cubes (side edge $d$) with different size ratios. Top: $\sigma/d$=1.0 (equal inradius), center: $\sigma/d$=1.23 (equal ODPs), and bottom: $\sigma/d$=1.732 (equal circumradius). S = sphere rich solid, C = cube rich solid, I = isotropic phase; $P^*=Pd^3/\epsilon$. Data for $\sigma/d$=1.23 is from \cite{Escobedo2014}. Orientation correlation plots are shown for the cubes in the 3 phases occurring at the eutectic pressure.}  \label{fgr:spcu}  \end{figure}

 To discriminate the role on mixture phase behavior of particles with equal ODP vs. particles with equal circumradius, the components should not both be {\em round-shaped} but one of them have high asphericity. For contrast, we simulated  mixtures of spheres and cubes. Spheres can be seen as the limiting case of a rounded polyhedra, whose FCC solid can also be taken to be a rotator if a minimal shape anisotropy is assumed \cite{Bolhuis1997}. Cubes can be seen as the limiting member of the truncated cube family having minimal truncation and high asphericity, whose solid phase is no longer a rotator\cite{Agarwal2011}. Figure \ref{fgr:spcu} shows the phase diagrams traced using a Gibbs-Duhem integration method \cite{SM, Escobedo2014}. Results are shown for 3 choices of the sphere diameter $\sigma$ to cube edge $d$ ratios: 1 (equal inradius), 1.23 (equal ODPs), and 1.732 (equal circumradius). Equal circumradii leads to minimal mutual solid solubility and an almost non-existent MRM region. In contrast, equal ODPs  lead to maximized {\em mutual} solid miscibility with both a large region where spheres dissolve in the cube-rich solid (C region) and   a large MRM region  where cubes dissolve in the sphere-rich solid (S region in gray) . In that latter  MRM, the orientation scatterplot (Fig. \ref{fgr:spcu} ) reveals that cubes form a restricted rotator  where they lack orientational order but can't adopt certain orientations. Such orientational correlations (e.g., see Figure \ref{fgr:mainfig}) depend on the shape and size of the particle relative to those of the cage where it rattles.\cite{Thapar2013}               \par The above analysis suggests that the ODP is a more generally predictive parameter of solid-phase miscibility of two shapes (beyond rotator mesophases). The ODP can be seen as marking the turning point where packing entropy takes over as the dominant entropic force determining the structure of the system. Accordingly, if the components have the same ODP, their tendencies to order will be comparable (i.e., synchronized) at any pressure above this ODP. In Fig. 1, the components have synchronized their rotator mesophases along the scale of the thermodynamic field driving the phase transitions (i.e., pressure). Indeed, for A+B mixtures, if ODP$_A$ $ \ll$ ODP$_B$ then for ODP$_A< P<  $ODP$_B$ particles B will have a strong preference for the isotropic state, while for $P >$ ODP$_B$ where both favor ordered states, particles A would be much more compressed than those of B and prone to form a separate A-rich dense solid. If one considers the pure components and that $\mu^{*}=\int_{\textup{\tiny 0}}^{\textup{\tiny ODP}} (Z-1)/P dP$ is the residual chemical potential of the isotropic phase in coexistence with the ordered phase ($Z$ is the compressibility factor), then for hard-core systems whose isotropic branches of the equation of state are similar (see Figure 1 in \cite{SM}), having equal ODPs translates into pure mesophases that at the same pressure also have comparable chemical potentials and (neglecting the typically small $\Delta PV$ terms) similar entropies. If rotational entropies are also comparable (as in rotator phases), equality of ODPs then approximately translates into pure mesophases of A and B where each particle experiences a similar packing entropy or {\em free volume}: a likely helpful condition for co-assembly.  \par As a final test of the equal-ODP rule, we simulated a {\it ternary} equimolar mixture of COs, TC4s and TOs at ODP-matching ratios, and found that the ternary MRM is also stable (with $\Delta P_m^* \approx 3.6$; see \cite{SM}). Of course, equality of ODPs is not sufficient to ensure high solid-phase compatibility; similarity in the type of ordered structure is also important as with the rotator mesophase in the COs, TC4s, and TOs; in this context, the sphere-cube system provides a counter example where solid miscibility over all compositions is precluded by the different pure-component solid behavior.      \par Recent work from Van Anders et. al. \cite{VanAnders2013,VanAnders2014} described the assembly of anisotropic particles as driven by an  entropic bonding  arising from `patches' that is quantifiable via a potential of mean force and torque (PMFT)(akin to enthalpic interactions). As the MRM is compressed and the patches get closer, any PMFT difference between like and dislike particles become more accentuated, making the mixed state less entropically favorable. This effect is connected with the changes in local composition discussed before regarding Figure \ref{fgr:cotoxplots}-b: like-particle contacts are favored with increasing density as though an effective attraction (repulsion) acts between the like (unlike) particle types. Eventually, the entropic cost at higher densities overpowers the mixing entropy leading to phase separation into two solids (this analysis does not apply to  tessellating polyhedral compounds \cite{Khadilkar2012}).      \par Beyond polyhedral particles, binary mixtures of rigid rods (of diameters $D_1$ and $D_2$ and lengths $L_1$ and $L_2$) with ODPs associated with isotropic-nematic transitions provide further insights. Simulation \cite{Dijkstra1997} and Onsager's theory {\cite{Hemmer1999} have shown that rods sufficiently dissimilar in length and$/$or diameter phase separate into two nematic phases at high pressures (a sign of incompatibility). However, `symmetric' mixtures \cite{Sear1996} where $L_2/L_1=(D_2/D_1)^{-\frac{1}{2}}$ so that pure components have the same excluded volume  and hence identical ODPs, tend to lie well inside the predicted one-nematic phase domain (see Figure 3 of Ref. \cite{Hemmer1999}), with equimolar mixtures having components with the same extent of orientational order (a sign of maximal compatibility) and ordering at pressures below the pure-component ODPs\cite{Sear1996}. Further,  novel  biaxial nematic phases have also been predicted for equal-ODP (symmetric) blends of rod-like and plate-like ellipsoids \cite{Camp1996,Camp1997,Varga2002}.  Note that in these examples and our simulated systems, ODP equality is not a prescription that guarantees {\em full} mesophase miscibility (which could only happen when particle shapes and pure-component behaviors are not too disparate); instead, it provides a guideline for conditions that favor miscibility (even if only a partial one, as for the cube-sphere example of Fig. 3).   \par In summary,  we find that by choosing size ratios that synchronize the onset of the plastic crystals in the pure components of a  mixture,  fully mixed mesophases are favored despite incompatibilities in the lattice structure of the pure component crystals. A vast array of applications \cite{Colvin2001, Huynh2002,Gur2005,Leschkies2009,Lim2012,VonFreymann2010} will benefit from new routes to create nanoparticle superstructures. Just like liquid-crystal phases have found widespread applications as switches and sensors, rotator phases may also find applications involving the external control of their rotational state. Since components can have different chemistries, the ability to produce rotator phases of any composition should add to this potential.   \par This work was supported by the U.S. National Science Foundation, Grant No. CBET 1402117. The authors also thank Dr. U. Agarwal for useful exchanges.  \bibliography{wms}   \end{document}